\documentclass[conference]{IEEEtran}
\IEEEoverridecommandlockouts
\usepackage{cite}
\usepackage{varioref}
\usepackage{amsmath,amssymb,amsfonts}
\usepackage{algorithmic}
\usepackage{graphicx}
\usepackage{textcomp}
\usepackage{xcolor}
\usepackage[switch]{lineno}
\usepackage{url,hyperref}
\usepackage{adjustbox}
\usepackage{float}
\usepackage[T1]{fontenc}
\usepackage[utf8]{inputenc}
\usepackage{tikz}
\usepackage{pifont}
\usepackage{scalerel}
\usepackage{ifthen}
\usepackage{pdfpages}
\usepackage{amsmath,amssymb,amsfonts}
\usepackage{algorithmic}
\usepackage{graphicx}
\usepackage{textcomp}
\usepackage{xcolor}
\usepackage[switch]{lineno}
\usepackage{url,hyperref}
\usepackage{adjustbox}
\usepackage{float}
\usepackage[T1]{fontenc}
\usepackage[utf8]{inputenc}
\usepackage{tikz}
\usepackage{pifont}
\usepackage{scalerel}
\usepackage{ifthen}
\usepackage{tabularx}
\usepackage{subcaption}
\usepackage{wasysym}
\usepackage{marvosym}
\usepackage{booktabs}
\usepackage{multicol}
\usepackage{multirow}
\usepackage{ragged2e}
\usepackage{xspace}

\usepackage[linesnumbered,ruled,vlined]{algorithm2e}

\usepackage{fancyhdr} 
\usepackage{lastpage}
\usepackage{authblk}
\def\BibTeX{{\rm B\kern-.05em{\sc i\kern-.025em b}\kern-.08em
    T\kern-.1667em\lower.7ex\hbox{E}\kern-.125emX}}

\SetCommentSty{mycommfont}

\SetKwInput{KwInput}{Input}                % Set the Input
\SetKwInput{KwOutput}{Output}              % set the Output
\SetKwInput{KwFunction}{Function}              % set the Function

\begin{document}

% \linenumbers

% \title{Towards a Privacy-preserving Distributed Cloud Service for Preprocessing Very Large Medical Images  \\
% \title{Privacy-, Budget-, and Time-Constrained Service Allocation Optimization in Hybrid Clouds for Large Medical Image Processing \\

% \title{Privacy-, Budget-, and Deadline-Aware Service Allocation Optimization for Large Medical Image Processing in Hybrid Clouds \\

\title{Towards Privacy-, Budget-, and Deadline-Aware Service Optimization for Large Medical Image Processing across Hybrid Clouds \\

\thanks{This work has been funded by the European Union project CLARIFY (860627), partially funded by $\text{ENVRI}^\text{FAIR}$ (824068), BlueCloud-2026 (101094227) and LifeWatch ERIC.}
}
% \author{Michael~Shell,~\IEEEmembership{Member,~IEEE,}
%         John~Doe,~\IEEEmembership{Fellow,~OSA,}
%         and~Jane~Doe,~\IEEEmembership{Life~Fellow,~IEEE}% <-this % stops a space
% \IEEEcompsocitemizethanks{\IEEEcompsocthanksitem M. Shell was with the Department
% of Electrical and Computer Engineering, Georgia Institute of Technology, Atlanta,
% GA, 30332.\protect\\
% % note need leading \protect in front of \\ to get a newline within \thanks as
% % \\ is fragile and will error, could use \hfil\break instead.
% E-mail: see http://www.michaelshell.org/contact.html
% \IEEEcompsocthanksitem J. Doe and J. Doe are with Anonymous University.}% <-this % stops a space
% \thanks{Manuscript received April 19, 2005; revised August 26, 2015.}}

% \author{\name{Yuandou Wang$^{1}$, Neel Kanwal$^{2}$, Kjersti Engan$^{2}$, Chunming Rong$^{2}$, Paola Grosso$^{1}$, Zhiming Zhao$^{1,3}$}\\
\author[1]{Yuandou Wang}
\affil[1]{Multiscale Networked Systems, University of Amsterdam, The Netherlands}
% \email{}
\author[2]{Neel Kanwal}
\affil[2]{Department of Electrical Engineering and Computer Science, University of Stavanger, Norway}
\author[2]{Kjersti Engan}
% \affil[2]{Department of Electrical Engineering and Computer Science, University of Stavanger, Norway}
\author[2]{Chunming Rong}
% \affil[2]{Department of Electrical Engineering and Computer Science, University of Stavanger, Norway}
\author[1]{Paola Grosso}
% \affil[1]{Multiscale Networked Systems, University of Amsterdam, The Netherlands}
\author[1]{Zhiming Zhao}
% \affil[1]{Multiscale Networked Systems, University of Amsterdam, The Netherlands}

% % \affiliation{$^1$Multiscale Networked Systems, University of Amsterdam, The Netherlands\\
% $^2$Department of Electrical Engineering and Computer Science, University of Stavanger, Norway\\
% $^3$LifeWatch ERIC Virtual Lab and Innovation Center (VLIC), Amsterdam, The Netherlands\\
% Email: \{y.wang8, p.grosso, z.zhao\}@uva.nl; \{neel.kanwal, kjersti.engan, chunming.rong\}@uis.no\\
% }
% }

\maketitle

% \textbf{This is on \date{\today}}

\begin{abstract}
Efficiently processing medical images, such as whole slide images in digital pathology, is essential for timely diagnosing high-risk diseases. However, this demands advanced computing infrastructure, e.g., GPU servers for deep learning inferencing, and local processing is time-consuming and costly. Besides, privacy concerns further complicate the employment of remote cloud infrastructures. While previous research has explored privacy and security-aware workflow scheduling in hybrid clouds for distributed processing, privacy-preserving data splitting, optimizing the service allocation of outsourcing computation on split data to the cloud, and privacy evaluation for large medical images still need to be addressed. This study focuses on tailoring a virtual infrastructure within a hybrid cloud environment and scheduling the image processing services while preserving privacy. We aim to minimize the use of untrusted nodes, lower monetary costs, and reduce execution time under privacy, budget, and deadline requirements. We consider a two-phase solution and develop 1) a privacy-preserving data splitting algorithm and 2) a greedy Pareto front-based algorithm for optimizing the service allocation. We conducted experiments with real and simulated data to validate and compare our method with a baseline. The results show that our privacy mechanism design outperforms the baseline regarding the average lower band on individual privacy and information gain for privacy evaluation. In addition, our approach can obtain various Pareto optimal-based allocations with users' preferences on the maximum number of untrusted nodes, budget, and time threshold. Our solutions often dominate the baseline's solution and are superior on a tight budget. Specifically, our approach has been ahead of baseline, up to  85.2\% and 6.8\% in terms of the total financial and time costs, respectively. 

\end{abstract}


\section{Introduction}

Medical data collected by hospitals and clinics are usually stored and processed in individual infrastructures due to privacy concerns. 
In recent years, deep learning (DL) technologies have been widely used in many diagnostic and prognostic tasks, for example, in digital pathology (DP)~\cite{kanwal2022devil,kanwal2023vision}, dealing with large whole slide images (WSI) produced by microscopy scanners. This is because DL can potentially improve the accuracy of automated diagnosis systems and save clinicians' time by providing second opinions. Clinicians or researchers might want to use DL-based methods over thousands of images, but WSIs are gigapixel images and one WSI may size up to 20GB in uncompressed form~\cite{kanwal2022devil}, thus it becomes computationally heavy. For instance, finding a segmentation map for different tissue types in a WSI using deep neural network (DNN)-based features may take a few hours, and running the DNN models for datasets like TCGA\footnote{The Cancer Genome Atlas Program (TCGA). \url{https://www.cancer.gov/ccg/research/genome-sequencing/tcga}. Retrieved 4 October 2023.} (with over 30,000 WSIs from 25 cancer types) may take several years to conclude using a single GPU machine~\cite{kurc2015scalable}. Hospitals or individual research institutes usually do not have an inexhaustible infrastructure that supports complicated computation and it is expensive for a hospital to conduct such computation in their local machines.

Cloud computing offers ample and highly scalable storage and computational resources with various pricing models and ubiquitous access through virtualization technology. Many companies, institutes, or organizations are outsourcing some of their information technology and data processing to the cloud~\cite{domingo2019privacy}. Nevertheless, large medical image processing is accompanied by large challenges: 

\begin{enumerate}
    \item \emph{Privacy concerns}. Medical images, such as WSIs, usually contain sensitive personal information. Many medical data custodians are reluctant to export their sensitive data over public cloud computing due to the sensitive nature of the data, privacy concerns with upgraded legal data protection requirements (e.g., GDPR\footnote{GDPR, applicable as of May 25, 2018. \url{https://gdpr-info.eu/}} and HIPAA\footnote{Public Law 104-191 HIPAA, approved on August 21, 1996. \url{https://www.govinfo.gov/app/details/PLAW-104publ191}}), together with security concerns on frequent data breaches mainly centered around the public cloud service providers (CSPs)~\cite{domingo2019privacy,truong2021privacy}. Not only can the CSPs access, use, or even sell the data outsourced by their customers, but the service might also suffer from attacks or emerging threats that can lead to data breaches and compromise data confidentiality.
    \item \emph{The requirement of privacy-preserving and cost-effective solutions}. Running DNN models is resource-hungry, and cloud customers could be billed with a ``pay-as-you-go'' pricing model based on their actual resource usage (e.g., GPU services) over a specific period, such as an hour or minute. Introducing data and task parallelism for distributed processing over clouds is common to reduce time and cost~\cite{wang2023towards}. However, privacy concerns further complicate the way of distributed processing. Consequently, customizing a virtual infrastructure and scheduling the image processing workflow under privacy, budget, and time requirements are highly important and must be addressed. 
\end{enumerate}

In recent years, hybrid cloud (HC) has received much attention since it provides the best of the private and public worlds, combining the economies and efficiencies of public cloud computing with the security and control of private cloud computing~\cite{mazhelis2012economic, weinman2016hybrid,aryotejo2018hybrid}. 
Several existing works have studied privacy and security-aware workflow scheduling problems for distributed processing in HCs~\cite{wen2020scheduling, sharif2016privacy, zhou2019privacy,lei2022privacy}. However, most solutions focus on one or two objectives optimization (e.g., the execution time and monetary cost reductions). In those solutions, there are no explicit key metrics for privacy evaluation in the deployment and execution of applications. The vulnerabilities and uncertainties of the system are implicitly described in the data movement or task execution in the whole workflow execution procedure, which may not solve the data privacy issues in the distributed data processing for different purposes. This paper aims to overcome those limits by providing a two-phase solution to protect data privacy during the distributed processing service, minimize the number of untrusted processing nodes, reduce the financial cost, and decrease the execution time over an HC. Those constraints and optimization objectives are formally stated as a multi-objective optimization (MOO) problem.

\textbf{Our novel contributions} 
% In this paper, we formulate the problem statement as a MOO workflow problem and propose a two-phase solution for problem-solving that will be integrated into a collaborative virtual research environment based on the works~\cite{wang2022scaling, zhao2022notebook,wang2023towards}. To validate our methodology, we study a use case of medical image processing from the DP domain and conduct experiments on real and simulated data sets. 
 can be summarized as follows:
\begin{itemize}
    % \item We formulate the complicated problem as a MOO problem, aiming to customize a virtual infrastructure over an HC and schedule the big image processing workflow in a privacy-preserving and cost-effective manner. % a new research problem

    \item We design and implement a privacy-preserving data splitting algorithm for large medical images to not only introduce data parallelism but also achieve the goal of data protection for distributed processing. This is a critical prerequisite for privacy-preserving workflow scheduling in the HC environment.  % a new privacy-preserving algorithm for image splitting

    \item We develop a greedy Pareto front-based approach to seek 3-dimensional (3D) Pareto optimal solutions of the MOO problem, enabling to improve the algorithm's efficiency with less search space.  % a new service deployment planning algorithm for application deployment and execution
    
    \item We demonstrate how to analyze, quantify, compare and understand our algorithms by conducting experiments based on a use case from the DP domain. 
    % We validate the feasibility of our method using real and simulated data.
    % The results show that our approach performs better than the baseline method regarding the average individual privacy at the lower bound and average information gain and can be superior to the baseline regarding very tight budget cases.   % experimentation on a real-world use case from digital pathology domain
\end{itemize}

The remainder of this paper is structured as follows. Section~\ref{sec:related} presents related work on cryptographic and non-cryptographic techniques to data processing, and privacy-aware workflow scheduling in HCs. Section~\ref{sec:preliminar} provides related concepts used for the methodology and evaluation. In Section~\ref{sec:system}, we propose our system model and formulate the problem statement. Section~\ref{sec:method} illustrates our proposed two-step approach for problem-solving. Section~\ref{sec:experiments} details the experimentation and Section~\ref{sec:results} discusses the results. Finally, we conclude our work in Section~\ref{sec:conclusion}.

\section{Related Works}\label{sec:related}

% The increasing volume of sensitive medical data being harvested by data providers or custodians makes it necessary to use the cloud to store the data and perform calculations. Also, the wide adoption of DL approaches in healthcare has increased concerns about the privacy implications of DL models applied to information-sensitive data. 

This paper involves a solution to privacy and security-aware workflow scheduling over HCs for sensitive data based on multiple disciplines. Therefore, this section reviews the state of the art on privacy-preserving techniques to data processing and existing works for privacy-aware MOO problems over HCs.

% This section reviews the state of the art on privacy-preserving distributed processing on sensitive data and existing works for privacy-aware MOO problems over HCs.

% \textcolor{blue}{Hints: There are scenarios in which it is beneficial to split a large volume of data to store and process the split data on cloud premises. --> gap analysis}

% In terms of security, 

% Regarding privacy, 

% Privacy is even more challenging than security because it must also hold with respect to (public and, therefore, untrusted) CSPs~\cite{domingo2019privacy}.

\subsection{Privacy-preserving Techniques to Data Processing}
In the past few decades, various privacy-preserving techniques have been developed to protect sensitive data from unauthorized or untrusted CSPs, broadly categorized into two types, namely, cryptographic and non-cryptographic~\cite{domingo2019privacy, Singh2015}:

\subsubsection{Cryptographic Techniques}
Cryptographic techniques provide robust mechanisms to preserve privacy by transforming data into directly unreadable form. Approaches based on differential privacy~\cite{dwork2014algorithmic}, secure multiparty computation (MPC)~\cite{cramer2015secure}, homomorphic encryption (HE)~\cite{rivest1978method, gentry2009fully}, attribute access control~\cite{goyal2006attribute}, searchable encryption~\cite{zhang2017searchable}, and lightweight cryptographic procedures~\cite{zhou2017security}, are a few examples for cryptographic (or encryption) techniques which can provide good privacy~\cite{chamikara2021privacy}. However, a critical shortcoming of such approaches is the high computational and /or communication complexities. 

Differential privacy is a theoretical framework that assumes an extreme scenario where all nodes in the network are corrupted ($k_i=n-1$) except for node $i$. Nevertheless, it has a trade-off between privacy and utility which has been proved in~\cite{li2021privacy} in several examples, such as distributed average consensus and privacy-preserving distributed average consensus. It is argued that approaches based on differential privacy are information-theoretically possible but computationally intractable while maintaining a non-constant error increase, especially for data with a considerable size~\cite{vadhan2017complexity}. Because of the high noise levels imposed by differential privacy approaches, it might not be the optimal choice for big data privacy~\cite{chamikara2021privacy}.

Even though MPC-based approaches can obtain full utility without compromising privacy, it is hard to generalize to problems other than distributed average consensus, and the robustness over $n-1$ corrupted nodes is no longer achievable~\cite{li2021privacy}. Besides, Boyle \textit{et al.}~\cite{boyle2018bottleneck} demonstrates the existence of $n$-party functions with $k$ bits of input for each part that has bottleneck complexity $\Theta(nk)$, which is defined as the maximum communication required by any party. In addition, MPC has to exchange information between nodes in the processing and could, hence, reveal some information about the private data~\cite{li2021privacy}, which may increase the way of distributed processing. HE-based approaches draws significant attention as a privacy-preserving way for cloud computing because it allows users to compute the encrypted data (called ciphertexts) without decrypting it~\cite{chakarov2019evaluation}, but the computation on ciphertext is arithmetically heavy, especially HE multiplication. For example, the execution time for computation on encrypted data (ciphertext) increases from 100s to 10,000s of times compared to that on native, unencrypted messages~\cite{jung2021accelerating}. Thus, implementing these techniques is rather expensive and impractical (especially for big data privacy) with respect to efficiency in distributed systems. 

\subsubsection{Non-cryptographic Techniques}
% Non-cryptographic techniques do not rely on encryption. 
Compared to cryptographic techniques, non-cryptographic ones such as data perturbation and data splitting approaches for big data privacy provide lightweight and efficient implementation; therefore, they are more appropriately fitting for outsourcing data in the cloud. 
Some of the common manifold non-cryptographic techniques are access-controls~\cite{sandhu1994access}, pseudonymization~\cite{neubauer2011methodology}, data anonymization~\cite{bayardo2005data}, and privacy-preserving data splitting~\cite{domingo2019privacy,farras2021privacy}. 

Data anonymization methods, currently the most widely used utility-based approach~\cite{kaissis2020secure}, rely on obscuring or removing sensitive information from medical data. There are two types of data anonymization considering masking methods~\cite{vijayarani2011efficient}: 1) non-perturbative masking introduces partial suppression or reductions in certain values which does not alter the truthfulness of the data but it reduces their accuracy, and 2) perturbative masking, which perturb (or alter) original values to limit disclosure risk by creating uncertainty around the true values. The latter is better than the former since it may preserve statistical properties for big data privacy, and thus, we investigate more on perturbative masking~\cite{domingo2019privacy, li2021privacy}. The best-known perturbative masking methods include noise addition~\cite{brand2002microdata}, random-value perturbation~\cite{kargupta2005random}, rank swapping~\cite{rodriguez2019utility}, data shuffling~\cite{muralidhar2006data}, and microaggregation~\cite{domingo2005ordinal} methods. Due to the modifications, perturbative methods often suffer from utility. Various privacy models are introduced~\cite{chamikara2021privacy} to address the utility-privacy trade-off, e.g., $k-$anonymity~\cite{samarati1998protecting}, entropy $l-$diversity~\cite{machanavajjhala2007diversity}, and $t-$closeness~\cite{li2006t}. However, these models may suffer from the composition attack~\cite{ganta2008composition}, foreground information~\cite{wong2011can}, and $p-$optimality attack~\cite{zhang2007information}. 

Data splitting, as another non-cryptographic approach, aims to protect data privacy by fragmenting sensitive data and storing the fragments in different locations so that individual parts do not disclose identities or confidential information~\cite{domingo2019privacy,farras2021privacy}. Sánchez \textit{et al.}~\cite{sanchez2017privacy} presented a semantic data splitting approach for data outsourced to the cloud. The authors assess the semantics of the data to be protected and the user's privacy requirement. The work of Farràs \textit{et al.}~\cite{ farras2021privacy} is related to the data splitting approach we propose in this paper. It also has the potential for seeing the data splitting problem as a graph-coloring problem (i.e., hypergraph-coloring problem) and outsource computation on split data, however it differs from our proposed method in important ways. It cannot be straightforwardly applied for large image splitting because it is based on plain textual documents instead of gigapixel medical images. In addition, the method of Farràs \textit{et al.}~\cite{ farras2021privacy} still has several limitations related to customizing a virtual infrastructure for distributed data processing. 

% This work focuses on handling a large dataset $\mathcal{D}$ of medical images (e.g., WSIs). 

\subsection{Privacy-aware Workflow Scheduling in HCs}
% The outsourcing computation on various tasks also introduces complexity towards developing a preferable cloud resource allocation plan that reduces monetary and time costs while guaranteeing privacy requirements. 
% Scheduling privacy-aware workflows has been increasingly studied under privacy or security requirements in recent years. 
% Regarding the workflow scheduling in hybrid clouds, there are two types of assumptions considering privacy: 1) an application workflow is entirely private, and 2) an application workflow is partially private~\cite{sharif2016privacy}. The first assumption is outside the scope of this work because if the whole workflow is private, it cannot be outsourced or executed anywhere except on trusted private clouds or on-premises servers. The second assumption is more complicated and receives much attention because of the mixed privacy requirements or constraints for sensitive data, and computational tasks on such data can be scheduled on either private or public cloud resources. Our work is based on the second assumption and addresses such issues. In the following, we review privacy-aware workflow scheduling in hybrid clouds. 
There is a scarcity of literature that formulates privacy-aware workflow scheduling as an optimization problem adhering to privacy and security requirements. Sharif \textit{et al.}~\cite{sharif2016privacy} presented a multi-level task partition method to minimize the overall execution cost of workflows in HCs while considering the trade-off between time and privacy. Three multiterminal-cut for privacy in hybrid clouds (MPHC) policies are introduced to preserve the privacy of tasks. 
% Wen \textit{et al.} proposed the Deploy on Federated Cloud Framework (DoFCF) that dynamically partition scientific workflows across federated cloud (public/private) data centers for minimizing the financial cost while meeting the security requirements and handling run-time failures~\cite{wen2016dynamically}.
Zhou \textit{et al.}~\cite{zhou2019privacy} introduced the special privacy requirements and formulated the process mapping problem in geo-distributed data centers as a constraint optimization problem, aiming at minimizing the overall communication cost function of the application while meeting data privacy constraints. However, such approaches are highly dependent on the application structure and heavily limited by the allocation map related to privacy privileges defined by authors.
Similarly, Lei \textit{et al.}~\cite{lei2022privacy} investigated how to effectively reduce monetary cost when deploying a workflow application in an HC under the deadline and privacy constraints. Wen \textit{et al.}~\cite{wen2020scheduling} proposed a multi-objective privacy-aware (MOPA) scheduling algorithm based on genetic algorithm (GA), which aims to obtain a set of Pareto trade-off solutions between execution makespan and monetary cost reductions while meeting the set of privacy protection constraints. 

The main weakness of all these solutions is that they are said to satisfy privacy requirements while reducing the cost or the execution time. However, they do not solve any privacy issues relevant to realistic applications. As a result, such workflow scheduling problem formulations and solutions still have the disadvantages of privacy protection and no explicit metrics for privacy evaluation, even though they had good strategies for the execution time and monetary cost.

\section{Background}\label{sec:preliminar}
% The proposed method uses privacy-preserving data splitting and Pareto MOO, in which we first protect the privacy of the data, and then optimize the distributed data processing with reference to multiple objectives. 

% This section provides the definitions and evaluation metrics used for 1) threat and adversary models and 2) the MOO problem and its solutions -- Pareto frontier. 

\subsection{Threat and Adversary Models}
In this work, we consider CSPs that are \textit{honest-but-curious}, in which CSPs honestly fulfill their role in the storage and processing services, but may inspect the information that users store or process~\cite{domingo2019privacy,farras2021privacy}. Adversary models are used to evaluate the robustness of the system under different security attacks~\cite{li2021privacy}. We consider two widely-used adversary models, i.e., passive and eavesdropping adversary. 

\subsubsection{Passive Adversary}
% The passive adversary is known as semi-honest or honest-but-curious threat~\cite{goldreich2009foundations}. It typically tries to collude a number of nodes to infer and reason sensitive information of the other nodes to gain insights without directly altering the data or system. These colluding nodes are called corrupted nodes $\mathcal{N}_c\in \mathcal{N}$, and the others are referred to as honest nodes $\mathcal{N}_h\in \mathcal{N}$. In a network graph, an edge is corrupted when there is one corrupted node at its ends. The passive adversary will know all the messages transmitted along such an edge.  
The passive adversary is known as a semi-honest or honest-but-curious threat~\cite{goldreich2009foundations}. It typically tries to collude with a number of nodes to infer and reason sensitive information of the other nodes to gain insights without directly altering the data or system. These colluding nodes are called corrupted nodes, and the others are referred to as honest nodes. In a network graph, an edge is corrupted when there is one corrupted node at its ends. The passive adversary will know all the messages transmitted along such an edge. 

\subsubsection{Eavesdropping Adversary}
The eavesdropping adversary works by secretly monitoring communication channels or data transmissions to capture sensitive information without the knowledge of the communication parties~\cite{pawar2011securing,li2021privacy}. 
% To protect against eavesdropping adversaries, all messages transmitted through securely encrypted channels are essential. However, the cost of channel encryption is computationally demanding; therefore, it is also an essential factor to consider when designing privacy-preserving algorithms~\cite{pawar2011securing,li2021privacy}. 
In practice, we can configure secure channel for communication, such as SSL/TLS certificates, so that we assume the transmitted messages can be protected from eavesdropping adversaries.

\subsection{Key Aspects for Privacy Evaluation}
% \subsection{Key aspects for privacy evaluation}
\label{sec: key aspects}
% There are considerable privacy-preserving algorithms, each of which is derived from a different context and has different metrics and assumptions. 
To evaluate algorithms in the context of privacy preservation, we focus on information-theoretic security metrics, such as the average increase in information entropy, output utility, and individual privacy, since they assume a stronger adversary and are more efficient concerning both communication and computational demands~\cite{sankar2013utility, lindell2000privacy, li2021privacy}. 

% \paragraph{Resistance against attacks}
% The essential purpose of attacks is to reason the original data from the collected data, such as naive inference (NI), known I/O, and ICA-based attacks. It can be obtained as the average and standard deviations of the difference between the original data and the perturbed data, as well as reconstructed data using ICA respectively~\cite{chamikara2020efficient}. 
% % Let $X$ be the original private data, $X^p$
% \begin{equation}
%     \text{std}(X, Z) = \Bigg\vert\sqrt{\frac{\sum(x_i-\mu_X)^2}{n}}-\sqrt{\frac{\sum(z_i-\mu_Z)^2}{n}}\Bigg\vert \label{NI}
% \end{equation}

% \begin{equation}
%     \text{std}(D, D^r)_{\text{ICA}} = \Bigg\vert\sqrt{\frac{\sum(d_i-\mu)^2}{n}}-\sqrt{\frac{\sum(d_i^r-\mu^r)^2}{n}}\Bigg\vert \label{ICA}
% \end{equation}
% where $\mu, \mu^p$, and $\mu^r$ are the mean values of the sets $D, D^p$, and $D^r$, respectively. 
% NI examines the difference between the original and perturbed data. ICA attacks were run on each perturbed data set to reconstruct data sets. A higher value represents a higher difference, implying higher resistance.

% \subsubsection{Information-theoretic Metrics}
\subsubsection{Definitions of Entropy and Mutual Information}

The Shannon entropy $H(\cdot)$ is an essential concept in the information theory of communication to quantify the uncertainty (or randomness) of an information source, which is often utilized to quantify the privacy (for the case of discrete random variables)~\cite{shannon1948mathematical,lin1991divergence}. 
% \paragraph{Entropy}
Let $x_j$ be the random value of variables $X$, $n$ denotes the number of data points in $X$, where $j\in n$. The entropy can be calculated as, 
\begin{equation}\label{entropy}
    H(X) = -\sum^{n}_{j=1}x_jlog_2x_j
\end{equation}

The mutual information $I(X;Z)$ between two random variables $X$ and $Z$ measures the dependence between $X$ and $Z$~\cite{kraskov2004estimating, li2021privacy}, quantifying the average reduction in uncertainty about $X$ that results from learning the value of $Z$, which can be stated formally as,
\begin{equation}
    I(X;Z)=H(X)-H(X|Z)
\end{equation}
where $H(X)$ is the entropy for data $X$ and $H(X|Z)$ is the conditional entropy for $X$ given $Z$. 

\subsubsection{Average Information Gain}
Let $Z_i \subset Z$ and $S_i\subset S$ be the $i^{th}$ encoded and private sub datasets, respectively. $n$ is the number of sub datasets. Information gain is calculated by comparing the entropy of the private data $S$ with the encoded data $Z$, which can be quantified as:
\begin{equation}\label{AIG}
    AIG(S;Z)=\frac{\sum^{n}_{i=1} H(Z_i)-H(S_i)}{n}, \quad i=1,...,n
\end{equation}
$AIG$ examines the amount of uncertainty reduced by altering the original dataset. A positive $AIG$ value indicates that identifying the original data $S$ using $Z$ is difficult. 

\subsubsection{Output Utility}
The output utility is to measure how close the estimated value $\hat{Y}_i$ of a privacy-preserving distributed processing algorithm is to its desired output $Y_i$, for each node $i \in \mathcal{N}$,
\begin{equation}
    u_i = I(Y_i;\hat{Y}_i), \quad \forall i \in \mathcal{N}
\end{equation}
where $0\leq u_i\leq I(Y_i; Y_i)$ and $u_i=I(Y_i, Y_i)$ indicates perfect output utility~\cite{li2021privacy}.  

\subsubsection{Individual Privacy}
Let $\mathcal{V}$ denotes the set of random variables containing all information collected by the adversaries throughout the whole processing. 
The individual privacy of honest node $i\in \mathcal{N}_h$ quantifies the amount of information about the private data $S_i$ learned by the adversaries, which is given by, 
\begin{equation}
    \rho_i = I(S_i; \mathcal{V}), \quad \forall i \in \mathcal{N}_h
\end{equation}
The smaller $\rho_i$, the more private the data is. The lower bound on individual privacy is defined by, 
\begin{equation}
    \rho_{i, min} = I(S_i; \{S_j, \hat{Y}_j\}_{j\in \mathcal{N}_c})
\end{equation}
in which, based on the definitions of adversary models, it can be concluded that adversaries always have knowledge of the private data $\{s_j\}_j = S_j\in \mathcal{N}_c$ and estimated outputs over corrupted nodes $\{\hat{y}_j\}_j = \hat{Y}_j\in \mathcal{N}_c$ where the maximum number of corrupted nodes denotes $\mathcal{N}-1$ out of $\mathcal{N}$, regardless of the algorithm used. Hence, $\{S_j, \hat{Y}_j\}_{j\in \mathcal{N}_c} \subseteq \mathcal{V}$ is the information known by the passive adversary model, we have $I(S_i; \{S_j, \hat{Y}_j\}_{j\in \mathcal{N}_c}) \leq \rho_i \leq I(S_i; S_i)$~\cite{li2021privacy}.

\subsection{Multi-Objective Optimization}
% what is MOP?
\subsubsection{Multi-objective Optimization Problem}
Multi-objective optimization aims to simultaneously optimize a group of often conflicting objectives. The general MOO problem is posed as follows~\cite{marler2004survey}:
\begin{align}
    \text{Minimize}_\mathbf{x} \quad \mathbf{F(x)} &= [f_1(\mathbf{x}), f_2(\mathbf{x}), ..., f_n(\mathbf{x})]^{T}\\
    \text{subject to, \quad} g_j(\mathbf{x}) &\leq 0, \quad j = 1,2,..., m\\
    h_l(\mathbf{x})&=0, \quad l = 1,2,...,e
\end{align}
where $n$ is the number of objective functions, $m$ is the number of inequality constraints, and $e$ is the number of equality constraints. $\mathbf{x}\in \mathbb{R}^n$ is a vector of decision variables and the set $\mathbf{X} \subseteq \mathbb{R}^n$ is the feasible set of $\mathbf{x}$, in which it depends on the $n$-dimensional application domain. $\mathbf{F(x)}$ is a vector of objective functions, $f_i(\mathbf{x})$, also called criteria, payoff functions, or value functions. 

\subsubsection{Feasible Solutions}
The gradient of $f_i(\mathbf{x})$ regarding $\mathbf{x}$ is formulated as $\nabla_\mathbf{x}f_i(\mathbf{x})\in \mathbb{R}^n$. The feasible set of decision vectors $\mathbf{X}$ is defined as the set $\{\mathbf{x}|g_j(\mathbf{x})\leq 0, j=1,2,...,m; h_l(\mathbf{x})=0, l=1,2,...,e\}$. Correspondingly, the feasible criterion space (also called the feasible cost space or the attainable set), $\mathbf{Z}$ is defined as the set $\{\mathbf{F}(\mathbf{x})|\mathbf{x}\in \mathbf{X}\}$. The feasibility here implies that no constraint is violated, while attainability means that a point in the criterion space maps to a point in the design space. Each point in the design space must map to a point in the criterion space; however, every data point in the criterion space does not necessarily correspond to a single data point $\mathbf{x}\in \mathbf{X}$. This is the main reason why there is typically no single global solution in MOO, and it is often necessary to determine a set of points that all fit a predetermined definition for an optimum. 
% MOO aims to find a set of solutions $\mathbf{x}_i^*$, where no solution is superior to another in all objectives. $\mathbf{x}_i^*\in \mathbf{X}$ is the point that minimizes the objective function $f_i(\mathbf{x})$, also called a feasible solution. 

\subsubsection{Pareto Front}
MOO aims to find a set of solutions $\mathbf{x}^*\in \mathbf{X}$, where no solution is superior to another in all objectives. Evaluating the solutions involves assessing the quality and trade-offs among all objectives. Pareto dominance represents a measure of the efficiency of the feasible solutions. From the mathematical point of view, the definition of the dominance between two candidate solutions can be expressed as $\mathbf{x_1}$ dominates $\mathbf{x_2}$ if $f_i(\mathbf{x_1}) \leq f_i(\mathbf{x_2})$, $\forall i=1,2,..., n$. Therefore, a solution $\mathbf{x}^*\in \mathbf{X}$ is called to be nondominated or Pareto optimal if and only if there does not exist any other point $\mathbf{x}\in \mathbf{X}$, such that $\mathbf{F}(\mathbf{x}) \leq \mathbf{F}(\mathbf{x^*})$ and $f_i(\mathbf{x})<f_i(\mathbf{x^*})$ for at least one function~\cite{brisset2015approaches}. Pareto front (or Pareto frontier) is the set of all such non-dominated Pareto optimal solutions.

\begin{figure*}[!ht]
    \centering
    \includegraphics[width=0.95\textwidth]{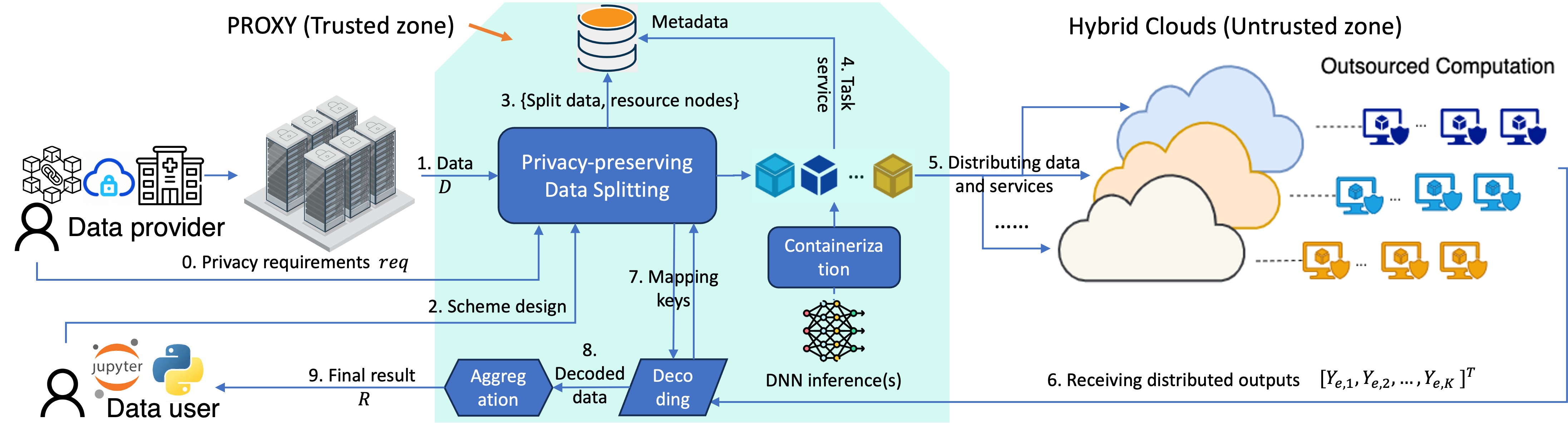}%
    \caption{Workflow of the proposed privacy-preserving distributed data processing with a very large volume of data.}
    \label{fig: distributed data processing}
    % \vspace*{-3mm}
\end{figure*}

\begin{figure}[!t]
    \centering
    \includegraphics[width=0.48\textwidth]{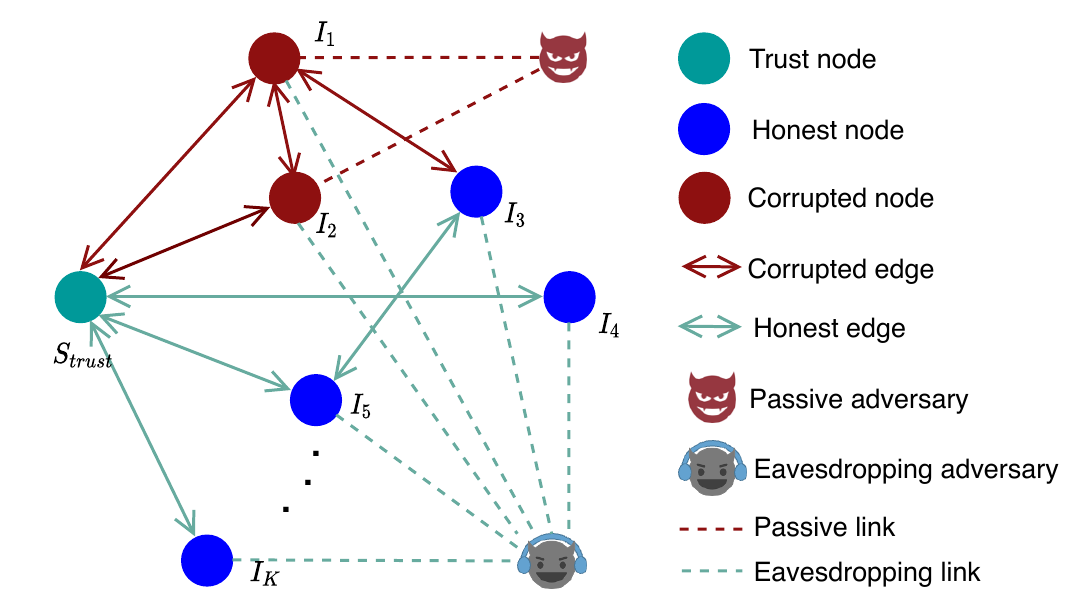}
    \caption{System setup and adversary models. }
    \label{fig: threat}
\end{figure}
\section{System Model and Problem Formulation}\label{sec:system}

% \includepdf[<options>]{<file>}

\subsection{System Model} \label{sec: sysm}

\subsubsection{System Architecture}
In order to customize a virtual infrastructure and schedule the privacy-preserving distributed processing, it is critical that we provide the user with a workflow for assessing the system. 
As depicted in Fig.~\ref{fig: distributed data processing}, our system model is composed of three main functionalities - viz \emph{Privacy-preserving Data Splitting} (steps 0$\to$1$\to$2$\to$3), \emph{Outsourcing Computation on Split Data}(steps 3$\to$4$\to$5$\to$6), and \emph{Data Aggregation} (steps 6$\to$7$\to$8$\to$9). The trusted zone 1) prepares the encoded data sets using our privacy-preserving data splitting, 2) containerizes the computational tasks (e.g., DNN inferences) as reusable fine-grained services (task containerization), 3) assigns split data and inference services to processing nodes, 4) aggregates the final results from decoded data. The untrusted zone consists of a set of available cloud instances in HCs, which may suffer from threats of adversaries. 

Specifically, Fig.~\ref{fig: threat} presents the system setup and adversary models. Let $\mathcal{G}=(\mathcal{N}, \mathcal{E})$ denotes a network from a provisioned infrastructure where $\mathcal{N} = \{S_{\text{trust}}, I_1, ..., I_K\}$ denotes the set of $K+1$ nodes and $\mathcal{E}\subseteq \mathcal{N}\times\mathcal{N}$ denotes the set of undirected edges, assuming it exists. We assume that $S_{\text{trust}}$ is a trusted zone that no adversary can interfere with. $\mathcal{I}=\{I_1, ..., I_K\}$ is assumed to be untrusted nodes where each node is modelled as a tuple $I_k = (\text{gpu}, \text{vcpu}, \text{ram}, \text{sto}, b, \text{loc}, p, \text{perf})_k$, equipped with the configurations about GPU devices, virtual CPU, RAM, local storage, network bandwidth, location, unit price, and performance score. These two zones cooperate, performing as a master-worker distributed computing model. The distributed data processing application can be defined as a tuple:
\begin{equation}
    \mathcal{A}=(\mathcal{M}, \varepsilon, \mathcal{D}, \mathcal{R},\mathcal{I}, req)   
\end{equation}
where $\mathcal{M}$ denotes a set of lighweighted interconnected pipeline steps which can also be seen as fine-grained \emph{services}. For instance, a \emph{splitting} service $m_{\text{splt}}$ processes the data stream produced by the source data set $\mathcal{D}$. A \emph{inference} service $m_{\text{inf}}$ is used for outsourcing the computation on split data. An \emph{aggregating} service $m_{\text{aggr}}$ is responsible for aggregating the final results $\mathcal{R}$. $\varepsilon$ indicates a set of \emph{data streams} $d_{u,i}$ flowing from an upstream service $m_u$ to a downstream service $m_i\in \mathcal{M}$. $\mathcal{I}$ denotes a set of \emph{cloud instances} over an HC and $req$ is a set of user requirements, e.g., privacy preservation, time and cost reduction.

% \begin{figure}[!ht]
%     \centering
%     \includegraphics[width=0.48\textwidth]{figs/WSI Pipelines [updates]-parallelism1.pdf}
%     \caption{A generic performance model considering Amdahl's law and our application for performance and scalability. In the classic program, some parts, such as parallel code fragment $f$, can be partitioned from the original program into reusable tasks run over $K$ processing nodes. Such division increases the parallelism and scalability but also introduces the communication overhead.}
%     \label{fig:paral}
% \end{figure}

\subsubsection{Privacy Guarantees}
We remove the privacy-sensitive information in the metadata from images to satisfy the privacy requirements. Then, we apply a privacy mechanism to generate distributed data fragments for outsourcing computation to available cloud instances. The mechanism is designed to hide critical attributes to prevent attackers from reconstructing the original image data considering the \emph{honest-but-curious} threat model. It protects privacy by avoiding exchanging the original data fragments among processing nodes. 
The output of the mechanism is a set of encoded split data sets $\{D_{e,1}, D_{e,2},..., D_{e, K}\}$. Each node $I_k\in \mathcal{I}$ stores and processes the split data $D_{e,k}$ with its private attributes $S_k$ and has its desired output $Y_{e,k}$. 
\begin{equation}
    f_{m_{\text{inf}}}: \mathbb{R}^K \mapsto \mathbb{R}^K, \quad Y_{e,k}=f_{m_{\text{inf}}}(D_{e,k}, I_k), \quad k\in K
\end{equation}
where $\mathbb{R}^K$ is the set of $K-$dimensional real vectors. $f_{m_{\text{inf}}}$ represents the outsource computation on split data with the service $m_{\text{inf}}$. Meanwhile, the total number of processing nodes should satisfy that, 
\begin{equation}
    \sum_{k=1}^{\mathcal{N_{I}}} x_k \geq \chi(\mathcal{D}), 
\end{equation}
where $x_k$ is a boolean value that if $m_{\text{inf}}$ and $D_{e,2}$ is mapping to $\mathcal{I}_k$, then $x_k$ equals 1; otherwise, 0. $\chi(\mathcal{D})$ is the smallest number of split data sets. Nevertheless, the ideal number of processing nodes in the distributed pipeline must still comply with the performance and cost requirements.

\subsubsection{Time Model}
% In the system, the cloud resource is resilient; thus, the time performance and scalability for the service deployment and execution can be improved by optimizing service allocation. 
The total time cost of the application can be comprised of the deployment time and execution time. 

\textbf{Deployment time.} 
The time for deploying one single task service $T_d$ typically consists of time cost for transmitting the data fragment $D_{e,k}$ and inference service $m_{\text{inf}}$ from the trusted zone to the target node (e.g., $I_k$) $T_f$, and installing the component $T_i$, respectively. Let bandwidth information for provisioned infrastructure as a set of $B=\{b_1, b_2, ..., b_K\}$, where $b_k$ denotes the bandwidth of $I_k$. The request of deployment service $r_k$ can be modelled as a tuple $r_k = (m_{\text{inf}}, D_{e,k})$. 
% The transmission time can be $T^{(k)}_f = \frac{size(r_k)}{b_k}$ if the service $(m_{\text{inf}}$ and data $D_{e,k}$ is assigned to $\mathcal{I}_k$. 
If $I_k$ responds to the deployment service request $r_k$, the deployment time is be given by, 
\begin{equation}
    T^{(k)}_d = T^{(k)}_f + T^{(k)}_i
\end{equation}
where $T^{(k)}_f=\frac{size(r_k)}{b_k}$ is measured by the size of the request $r_k$ and the network bandwidth between the trusted and target servers. In many cases, $T_f$ is much bigger than $T_i$~\cite{hu2017deadline}. Note that installation time prediction is not the focus of this work, and the size of $m_{\text{inf}}$ can be a constant, and hence we assume $T_i$ can be negligible and we only focus on the size of $D_{e,k}$.  

\textbf{Execution time.} 
% Based on the well-known Amdahl's law of strong scaling~\cite{amdahl1967validity}, performance can be theoretically deduced using a generic model for distributed system operation. Let $T_K$, as illustrated in Fig.~\ref{fig:paral}, be the time needed to execute the program over $K$ distributed nodes (i.e., the untrusted zone). For instance, $m_{\text{inf}}$ denotes a code fraction $f$ that can be distributed to untrusted zone. $m_{\text{splt}}$ and $m_{\text{aggr}}$ denote the sequential code fragments $1-f$ and $T^{(s)}_{\text{compt.}}$ is the execution time over the trusted server. 
% \begin{equation}
%     T_{\text{exe}} = T^{(s)}_{\text{compt.}} \times (1-f)+T^{(s,k)}_{\text{comm.}}+\max(T^{(k)}_{\text{compt.}}\times f)
% \end{equation}
The time for executing an application $T_{\text{exe}}$ typically consists of the time cost of on-site computation and transmitting the output to the trusted node. Let $T^{(k)}_{\text{comm.}}$ be the communication overhead between the trusted zone and distributed processing nodes during the execution period, the time cost over $I_k$ is given by,
\begin{equation}
    T^{(k)}_{\text{exe}} = T^{(k)}_{\text{compt.}} + T^{(k)}_{\text{comm.}}
\end{equation}
where $T^{(k)}_{\text{comm.}} = \frac{size(Y_{e,k})}{b_k}$ and the processing time mainly depends on the performance of $I_k$ and the complexity of the deployed service, i.e., $T^{(k)}_{\text{compt.}}= \text{perf}_k (r_k)$. Hence, the total time cost of the application over untrusted servers can be,
\begin{equation}
    % T_{\text{tot}} = T^{(s)}_{\text{compt.}}+\max(T^{(k)}_d+T^{(k)}_{\text{compt.}}+T^{(k)}_{\text{comm.}})
    T_{\text{tot}} = \max(T^{(k)}_d+T^{(k)}_{\text{compt.}}+T^{(k)}_{\text{comm.}})
\end{equation}
Note that the processing time $T^{(k)}_{\text{compt.}}$ typically far outweighs the communication time $T^{(k)}_{\text{comm.}}$ due to the far smaller size of output $Y_{e,k}$ than that of input $D_{e,k}$. Thus $T^{(k)}_{\text{comm.}}$ in the processing period is negligible. 

\subsubsection{Pricing Model}
% Commercial CSPs in HCs offer various pricing modes, such as on-demand pricing (well-known as the pay-as-you-go model), spot pricing, and reserved instance pricing. 
Let $p$ be the unit prices of $\mathcal{I}$, in which we consider commercial CSPs provide $\mathcal{I}_{c}$ with a pay-as-you-go pricing model and participating organizations offer $\mathcal{I}_{\mathcal{N_I}\setminus c}$ for free. 
% Suppose each commercial resource node follows a pay-as-you-go price model;
For untrusted nodes, the total monetary cost is, 
\begin{equation}
    \text{Cost} = \sum_{k=1}^{K} (T^{(k)}_{\text{d}}+T^{(k)}_{\text{exe}}) \times p_k
\end{equation}
where if $k\in \mathcal{I}_{c}$, then $p_k\in R^{+}$; otherwise, $p_k=0$. 

% Appendix Table~\ref{tab: infracloud} lists an example of the price model (\$/hour) over different configurations of cloud GPU virtual machines. 

% Note that CSPs offer various configurations of cloud instances, incl., the network bandwidth and compute capability, and pricing models. In this work, we focus on the transmission process (i.e., $T_f$) of deployment and the compute process ($T_{\text{compt.}}$) of execution and its financial cost. 

\subsection{Problem Formulation} \label{sec: problem_formulation}
We aim to reduce the total number of distributed processing nodes $f_1$, the total monetary cost $f_2$, and the maximum completion time over untrusted nodes $f_3$ under privacy, budget, and time constraints for the application deployment and execution, that requires MOO. 
% Assume that users have a set of requirements regarding privacy preservation, processing node usage, time cost, and monetary budget.
Therefore, our service allocation problem is formulated as follows:

\begin{align}
    \min f_1 =& \sum_{k=1}^{\mathcal{N_{I}}} x_k \label{eq:f1}\\
    \min f_2 =& \sum_{k=1}^{\mathcal{N_{I}}}(T^{(k)}_{\text{d}}+T^{(k)}_{\text{exe}}) \times p_k \times x_k \label{eq:f2}\\
    \begin{split}
        \min f_3 =& \max((T^{(k)}_d+T^{(k)}_{\text{compt.}}+T^{(k)}_{\text{comm.}}) \times x_k) \label{eq:f3}
    \end{split}\\
    \text{subject to, \quad \quad} 
    &\chi(\mathcal{D}) \leq f_1 \leq \text{\textit{Nodes}}, \\
    \quad\quad0 \leq &f_2 \leq \text{\textit{Budget}}, \\
    \quad0 < &f_3 \leq \text{\textit{Time Threshold}}, \\
    % T^{(k)}_d & = \\
    % T^{(s)}_{\text{compt.}} &= T(m_{\text{splt}}, m_{\text{aggr}}, S_{\text{trust}})\\
    % T^{(k)}_{\text{compt.}} &= T(m_{\text{inf}}, D_{e,k}, \mathcal{I}_k)\\
    % T^{(s,k)}_{\text{comm.}} &= TT(y_{e,k}, s, \mathcal{I}_k)\\
    x_k = &\begin{cases}
            1,  \text{if service $m_{\text{inf}}$ maps to } I_k,\\
            0,  \text{otherwise}.
            \end{cases}
\end{align}
where the total number of distributed processing nodes $f_1$ is limited to the privacy-preserving data splitting method and the maximum available resource \textit{nodes}. The total monetary cost $f_2$ is limited to the user \textit{budget}, and the maximum completion time $f_3$ is constrained by the user-defined \textit{time threshold}. 

% The communication overhead $T^{(s,k)}_{\text{comm.}}$ between the trusted zone (e.g., a trusted server $s$) and the untrusted zone (e.g., a distributed processing node $k$) is given by the transfer time sending distributed data fragments and receiving distributed outputs $[y_{e,1}, y_{e,2}, ..., y_{e, K}]^T$. It relies on the network bandwidth, protocols, and even the transferred data size. The higher bandwidth, the lower the transferred time, and the smaller the data size, so does the transferred time. In this work, we assume that the network bandwidth and protocols are high-performance; meanwhile, the size of split data sets $\{D_{e,1}, D_{e,2},..., D_{e, K}\}$ far outweighs that of distributed outputs. Thus $T^{(s,k)}_{\text{comm.}}(y_k)$ in the processing period has been negligible. 

\section{Our Solution: A Two-phase Approach}\label{sec:method}
This section explains our method to address the formulated problem with two phases: i) privacy mechanism design and ii) greedy Pareto frontier-based service allocation, respectively. 

\subsection{Phase I: Privacy mechanism design} \label{privacy mechanism design}
Phase I solves the privacy-preserving data splitting problem with image splitting and data perturbation techniques into two steps: (\textit{i}) First it greedily assigns (based on graph theory) the available colors to vertices $V$ with a random sequential (RS) strategy~\cite{kosowski2004classical}; 
(\textit{ii}) we adopt the random-value perturbation technique to hide critical information of image data sets.

\begin{algorithm}[!thb]
\DontPrintSemicolon
  \KwInput{WSI dataset $\mathcal{D}$, patch size $b$}
  % {Images $\mathcal{D}$, Patch size $s$, coordinates $A_x$, data set number $K$}
  \KwOutput{$K$, a new coordinate system $A_e$, and encoded data sets $D_{e,1},D_{e,2},..., D_{e,K}$.}

  % \tcc{Now this is creating patches from WSI}
  Patch set $D \gets$ create\_patches($\mathcal{D}$, $b$)

  % \tcc{Now splitting into data fragments}
  $G \gets$ create\_new\_coordinates($D$) 

  $\chi(\mathcal{D}), C \gets$ \textbf{graph\_coloring}($G$, `random\_sequential')
  
  $D_{x,1}, ...,D_{x,K} \gets$ divide $D_x$ into $\chi(\mathcal{D})$ sub data sets

  \For{$D_{x,k} \in D_{x,1}, ...,D_{x,K}$}{
    Data matrix $A_{x}$ of size ($n \times m$) $\gets D_{x,k} $

  % Initialize a random value $r$
  
  % $A_x$ $\gets$ random.shuffle($A$)
    
  % \tcc{Now this is normalization of each feature}
  $\bar{x} \gets$ mean($A_{x}$) 
  % \tcp*{Compute the mean}
  
  $\sigma \gets$ std($A_{x,i}$) 
  % \tcp*{Compute the standard deviation}
  
  $A_{x,c} \gets (A_x - \bar{x})/\sigma $ \tcp*{Normalize the data }

  \tcc{Compute the covariance matrix of $A_{x,c}$ } 
  $\Sigma \gets \frac{1}{n-1} A_{x,c}^T A_{x,c}$ 

  \tcc{Compute eigenvalues and eigenvectors}
  $\lambda, \vec{V} \gets$ \textbf{eig}($\Sigma$) 
  
  % \tcc{sort eigenvalues in descending order}
  $\lambda_{sort}, \vec{V}_{sort} \gets$ \textbf{sort\_eig}($\lambda$, $\vec{V}$)

  % \tcc{choose the top k eigenvectors}
  $\vec{V}_k \gets$ first $k$ columns of $\vec{V}_{sort}$ 

  $A_{e} \gets A_{x,c} \vec{V}_k$ 
  
  % \tcc{Transform to a new coordinate system with a random number $r$ to complete the coordinates}
  $D_{e,i} \gets \text{rename}(D_{x,i}, A_{e}, r)$ %\tcp*{r is a random value}
  }
  % \tcp*{Transform the data into the new coordinate system using the eigenvectors}
  
  \KwFunction{\textbf{sort\_eig}($\lambda$, $\vec{V}$)} 
  % \tcc{sort eigenvalues in descending order}
    \quad \quad$S \gets$ indices that sort $\lambda$ in a descending order
    
    \quad \quad$\lambda_{sort} \gets \lambda(S)$
    
    \quad \quad$\vec{V}_{sort} \gets \vec{V}(:, S)$
    
    \quad \quad\Return $\lambda_{sort}$, $\vec{V}_{sort}$

    \caption{Privacy mechanism design for large image splitting \CheckedBox}\label{alg: privacy method}
\end{algorithm}

At the trusted zone, we first apply a graph to abstract the positions of patches $D$ created by the code \textit{create\_patches}($\mathcal{D},b$), as described in Algorithm.~\ref{alg: privacy method}. Then, we transform the original coordinates with a 2D (x, y) axis into a new coordinate system (See pseudo code step 1$\to$2). 

Let $G=(V,E)$ be a graph extracted from $D$, in which each sliced patch is expressed as a vertex $\upsilon \in V$. Two vertices $\upsilon$ and $\mu$ of $V$ such that $(\upsilon, \mu) \in E$ are called to be adjacent. The RS strategy is a \textit{randomized} method, in which the vertices of the graph are randomly ordered for coloring assignment. The problem seeks to find the smallest number of possible colors, where $C$ is a set of colors, such that:
\begin{enumerate}
    \item $V^r$ := a random sequence of vertices of graph $G$;
    \item For every vertex $\upsilon \in V^r$, let $f(\upsilon)$ be the function that gives $\upsilon$ the smallest possible color;
    % $f(\upsilon)$ is a color from set $C$.
    \item For every edge $(\mu, \upsilon) \in E$, $f(\mu)$ and $f(\upsilon)$ are distinct colors.
\end{enumerate}
We can earn the minimum number of data fragments at this stage with graph coloring-based data splitting algorithm (step 3). However, the private attributes $(x, y)_\text{coord}$ of split data $D_{x,k}$, i.e., coordinates are plaintext; attackers may reconstruct the original data once these attributes are leaked. Hence, we inducted a random data perturbation-based method to hide such attributes. 

% The data perturbation procedure was applied to preserve the position on split data sets $\{D_{x,1}, D_{x,2},..., D_{x, K}\}$ where $D_{x}\in \mathcal{D}$, by inserting random noise. In our design, for each data set $D_{x,k} \subset D_x$, 
To do so, we extract $(x, y)_\text{coord}$ as a data matrix $A_x$ of size ($n \times m$). After normalization, we compute the covariance matrix of the normalized matrix $A_{x,c}$, and then computed the eigenvalues $\lambda$ and eigenvectors $\vec{V}$ so that we can get the top-k eigenvectors $\vec{V}_k$ to calculate $A_e$. Moreover, we transform the data into a new coordinate system and encoded into data sets $\{D_{e,1}, D_{e,2},..., D_{e, K}\}$. The pseudo code is illustrated from step 6 to step 15 of Algorithm.~\ref{alg: privacy method}. The proposed noise addition is efficient for numerical attributes in usability since noise is added to individual records independently. 
% It can preserve statistical features for large volume of data. 

\subsection{Phase II: Greedy Pareto Front based Service Allocation}
Phase II seeks Pareto optimal solutions to the MOO problem (See section~\ref{sec: problem_formulation}) since there are \textit{de facto} trade-offs among the total number $f_1$, the monetary cost $f_2$, and the time cost $f_3$ of processing nodes. For example, when diminishing the total number of processing nodes, in return, the time cost gains; these two factors together either decrease or increase the total monetary cost. The reverse is also true: once the processing node number is increased, the time cost will be reduced because of distributed parallel computing. Then we have to consider the change of cost, due to unit price. Such trade-offs are often illustrated graphically by a Pareto front. However, traditional methods could not be applied to the problem with large amounts of configurable items because of the significant search space. 
Thus, we propose a greedy Pareto front-based algorithm that allows for a more reasonable search space for performance improvement, as presented in Algorithm~\ref{alg: Paretosolutions}.
\begin{algorithm}[!bh]
\caption{Greedy Pareto Front based Service Allocation \CheckedBox}\label{alg: Paretosolutions}
\DontPrintSemicolon
    \KwInput{The set of workloads $\Theta$ and cloud instances $\mathcal{I}$}   
    
    \KwOutput{Pareto frontier $F_{p} = (F^S_{p}, F_{p}^N, F_{p}^T, F_{p}^C)$.}
    
    Initialize a feasible set $X$

    Initialize an empty Pareto frontier of solutions $F_{p}$  
    
    % \tcc{compute the Pareto front-based strategy, time, cost with privacy constraint $\chi(\mathcal{D})$}
    $S$, $T$, $C \gets $ \textbf{PF\_2D}($\Theta, \mathcal{I}, \chi(\mathcal{D})$) 
    
    $X \gets \textbf{combos}(S, T, C)$

    $F_{p} \gets \textbf{Paretoset\_3D}(X, \text{sMax=False)}$

    % \tcc{Now this is finding feasible solutions}

    % $\Theta, K, \mathcal{I}, p, T^{\text{berk}}_{\text{compt.}}$
    \KwFunction{\textbf{PF\_2D}($\Theta, \mathcal{I}, \chi(\mathcal{D})$)}
    
    \quad \quad Initialize $S$, $T$, $C$
    
    \quad \quad \For{$i = 0, 1,2,..., |\mathcal{D}|-1$}{

        \quad \quad index = $i \times \chi(\mathcal{D})$

        % \tcc{Find the Pareto front with parameters}
        \quad \quad $p^T, p^C, p^S  \gets $\textbf{Pareto\_2d}(index)

        \quad \quad add $p^T, p^C, p^S$ to $S, C, T$
        
        % \quad \quad $S.\text{append}(p^S)$ \tcp*{add optimal strategy}

        % \quad \quad $T.\text{append}(p^T)$ \tcp*{add optimal time}

        % \quad \quad $C.\text{append}(p^C)$ \tcp*{add optimal cost}
        
    }
    
    \quad \quad \Return $S$, $T$, $C$

    \KwFunction{\textbf{comos}($S$, $T$, $C$)}

    \quad \quad Nodes $\gets S_0^1S_1^1...S_{|\mathcal{D}|-1}^1 $ \tcp*{total nodes}

    \quad \quad Times $\gets T_0^1T_1^1...T_{|\mathcal{D}|-1}^1 $ \tcp*{time combinations}

    \quad \quad Costs $\gets C_0^1C_1^1...C_{|\mathcal{D}|-1}^1 $ \tcp*{cost combinations}

    \quad \quad calculate $f_1, f_2, f_3$ with equations~\eqref{eq:f1}, \eqref{eq:f2},~\eqref{eq:f3}
    
    \quad \quad \If{constraints}{ \tcp*{e.g., $f_1\leq \text{No}, f_2\leq budget, f_3\leq threshold$}

        \quad \quad $x \gets f_1, f_2, f_3 $
        
        \quad \quad add feasible solution $x$ into $X$
    }

    \quad \quad \Return $X$
    % \KwFunction{\textbf{PF\_3D}()}
    % \quad \quad    
\end{algorithm}

% Let $\Theta=(\theta_1, \theta_2,...,\theta_l)$, where $\theta_i=(m_{\text{inf}}, D_{e, i}) \in \Theta$ and $D_{e, i}\in \mathcal{D}$, as a set of workloads for outsourcing computation on split data (See in Fig.~\ref{fig: distributed data processing}). 
% Given the available resource nodes $\mathcal{I}$ and their relative performance measure based on their configurations. 
% We first initialized the time performance matrix $T_{\text{compt.}}$ of size ($l \times \mathcal{N_I}$) and the cost matrix $C$ and 

We first obtain the 2D Pareto front (See step 3) where we set the number of processing nodes equal $\chi(\mathcal{D})$ for each large image. Then, we calculate the combinatorial set with the obtained 2D Pareto optimal solutions (See step 4), which are the feasible (or candidate) solutions as input for seeking the 3D Pareto front (Step 5). We also set constraints regarding the maximum distributed processing nodes, time threshold, and cost budget to filter out unavailable solutions (See steps 16$\to$17$\to$18). And finally, we figure out the Pareto front based assignment for distributed service allocation.

\section{Experiments}\label{sec:experiments}
% In this section, we set up our experimental environment, illustrate the experimentation and evaluate the results for validation. 

\subsection{Experimental Setup}
% \subsubsection{Research Questions}
% The overall question we want to answer is \textit{how to customize a virtual infrastructure and schedule the service deployment and execution under privacy-preserving constraints while reducing its number of node s, time, and monetary cost?}

\subsubsection{Baseline}
To compare different approaches, we simulate each algorithm in a simple yet realistic way to ensure an accurate representation of their functionality. Our solution involves two phases, and the output of the first phase affects that of the second phase. Splitting the image equally to processing nodes is the simplest way; hence, we set it as the baseline, in which the patch-level dataset from each WSI is evenly split into sub-datasets. In the second phase, even though there are many efficient optimization algorithms such as non-dominated sorting genetic algorithms (NSGAII and NSGAIII), multi-objective particle swarm optimization (MOPSO), and multi-agent reinforcement learning (MARL)~\cite{wang2019multi} for MOO problem-solving, none of them has been applied to the same research question. Thus, comparing with different optimization algorithms in the second phase is out of the scope of this work, and we focus on data splitting and its influence on the final results.

\subsubsection{Use Case}
% \textbf{Data and Application}. 
We adopted a WSI preprocessing application using a DL-based method~\cite{kanwal2022quantifying} as our case study to test our methodology. Their method uses an ensemble of five MobileNet\_v3\_Large~\cite{mobilenet} architectures trained for different tasks with a central unit.
% Our work presents a privacy-preserving distributed data processing version of such a type of application. 

\textbf{Real data}. Table.~\ref{tab: data} presents the data characteristics, incl., digital slide size, the total number of patches, and data set size, from a public WSI data materials called TCGA\footnote{ \url{https://www.cancer.gov/ccg/research/genome-sequencing/tcga}} for experiments. Besides, the complexity of the DL inference model has been measured by parameters, MFLOPs, patch size, and memory. We have five such inference models. The total number of parameters and FLOPs account for approximately 17.65 million and 1.13 GFLOPs, respectively. 

% Table~\ref{tab: model computation cost} presents the complexity of the inference model measured by parameters, MFLOPs, patch size, and memory. We have five such inference models; thus, the total number of parameters and FLOPs account for approximately 17.65 million and 1.13GFLOPs, respectively. 
\begin{table}[!t]
    \caption{The dataset selected for testing. We randomly retrieved ten WSIs from NIH GDC Data Portal -- The Cancer Genome Atlas Program (TCGA) data repository (https://portal.gdc.cancer.gov/repository/). 
    % The selected data sets' sizes range from small to large, to allow the investigation of our method's performance on varying dimensions of data.
    }
    \label{tab: data}
    \centering  
    % \begin{tabular}{ccccc}
    % \toprule
    %     \textbf{No} & \textbf{Name} & \textbf{Slide size} & \textbf{Patches} & \textbf{Dataset size}\\
    %     \midrule
    %     I & A7-A0D9 & 204 MB & 7,740 & 1,03 GB \\
    %     II & A7-A3IZ & 171 MB & 11,637 & 1,49 GB \\
    %     III & E9-A1N3 & 281 MB & 20,589 & 2,44 GB \\
    %     IV & AC-A2FB & 180 MB & 20,469 & 2,65 GB  \\
    %     V & AC-A2FE & 311 MB & 4,040 & 500,1 MB  \\
    %     VI & B6-A0IJ & 286 MB & 13,002 & 1,49 GB  \\
    %     VII & E9-A1NE & 312 MB & 23,136 & 2,44 GB  \\
    %     VIII & E9-A1QZ & 551 MB & 40,412 & 4,16 GB \\
    %     IX & E9-A1R3 & 498 MB & 41,299 & 4,07 GB \\
    %     X & E9-A1R4 & 375 MB & 30,952 & 2,96 GB \\
    %     % \midrule
    % \bottomrule
    % \end{tabular}
    \begin{tabular}{cc}
    \toprule
        \textbf{No} & \textbf{Name}  \\
        \midrule
        I & TCGA-A7-A0D9-01Z \\
        II & TCGA-A7-A3IZ-01Z  \\
        III & TCGA-E9-A1N3-01Z  \\
        IV & TCGA-AC-A2FB-01Z \\
        V & TCGA-AC-A2FE-01Z \\
        VI & TCGA-B6-A0IJ-01Z \\
        VII & TCGA-E9-A1NE-01Z \\
        VIII & TCGA-E9-A1QZ-01Z \\
        IX & TCGA-E9-A1R3-01Z \\
        X & TCGA-E9-A1R4-01Z \\
        % \midrule
    \bottomrule
    \end{tabular}
\end{table}

\begin{figure*}[!ht]
    \centering
    \begin{minipage}[b][0.32\textheight][s]{0.30\textwidth}
        \centering
        \includegraphics[height=0.158\textheight,width=\textwidth]{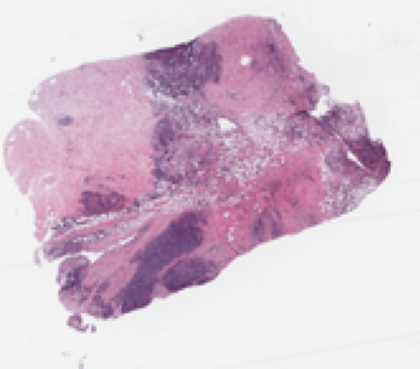}
        \vfill
        \includegraphics[height=0.158\textheight,width=\textwidth]{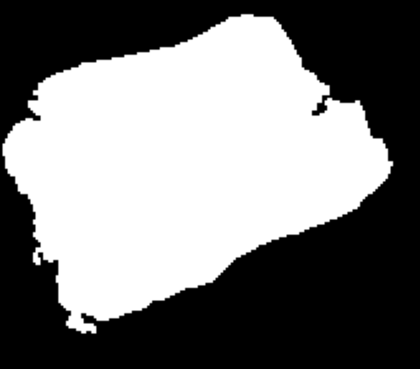}
    % \subcaption{thumbnail \& binary mask}
    \end{minipage}
    \includegraphics[height=0.32\textheight,width=.68\textwidth]{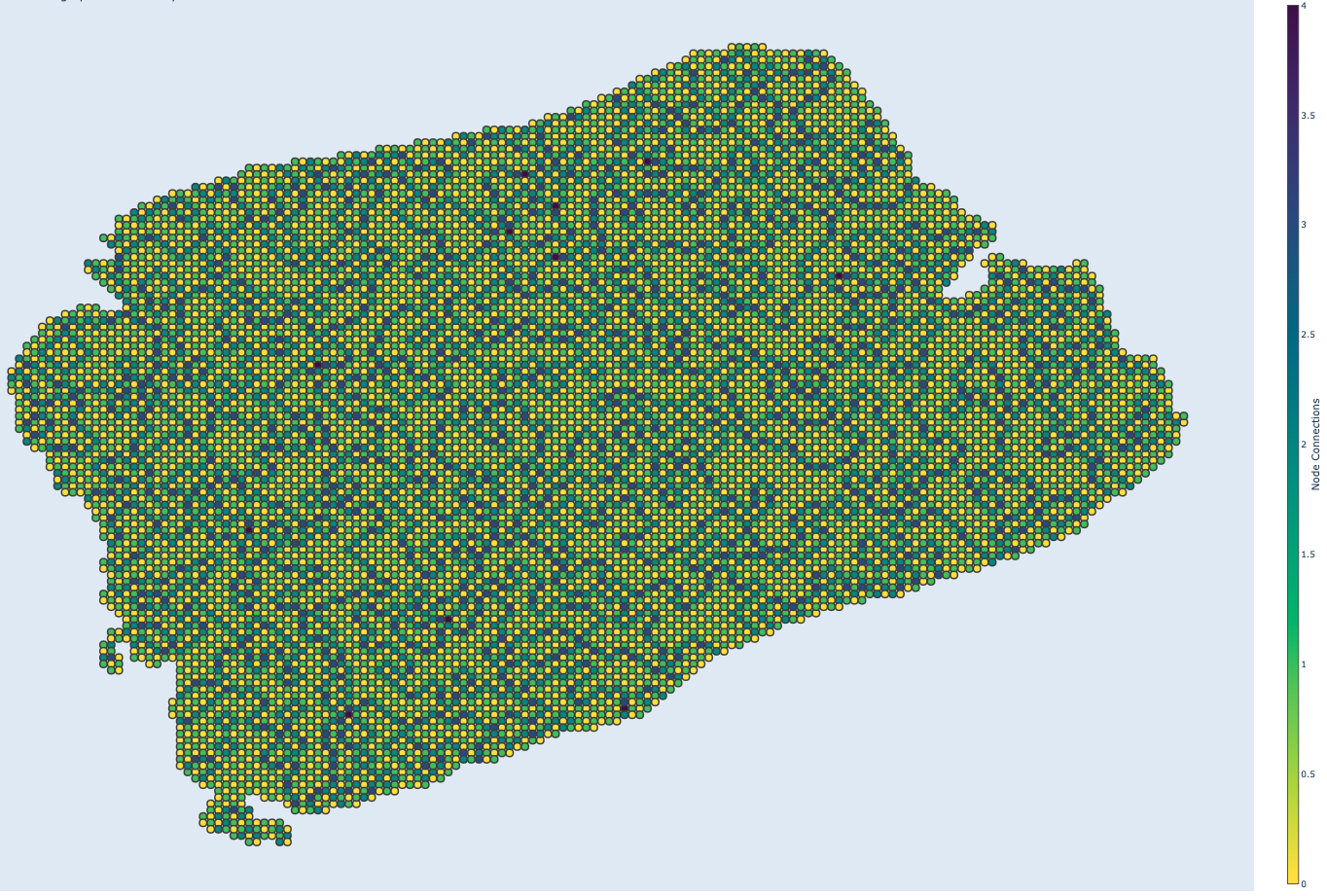}
    \caption{An overview of the visual results obtained by privacy-preserving data splitting. The three mini-pages of the figure, which are originally from the testing data No.II A7-A3IZ, are: thumbnail (top left), binary mask (bottom left) and data splitting results (right) on the basis of graph coloring.}\label{fig: ppdatasplit}
\end{figure*}
\begin{table*}[!tbh]
    \centering
    \caption{Comparisons of our approach with the baseline method in terms of the privacy-preserving data splitting.}
    \label{tab: privacyeval}
    \begin{tabular}{cccccccc}
    \toprule
        \multirow{2}*{\textbf{Method}} & \multirow{2}*{\textbf{No. Of sub-datasets}} & \multicolumn{2}{c}{\textbf{Output utility ($\uparrow$)}} & \multicolumn{2}{c}{\textbf{Lower bound on individual privacy ($\downarrow$)}} & \multicolumn{2}{c}{\textbf{Average information gain ($\uparrow$)}} \\ \cline{3-8}
        \textbf{} & \textbf{} & \textbf{$x\_coord$} & \textbf{$y\_coord$} & \textbf{$x\_coord$} & \textbf{$y\_coord$} & \textbf{$x\_coord$} & \textbf{$y\_coord$} \\
    \midrule
       Baseline  & 5  & 1.0$\pm$0.0 & 1.0$\pm$0.0 & 0.1714$\pm$0.0386 & 0.2876$\pm$0.0332 & 3.3099 & 3.7033\\
       Our approach & 5 & 1.0$\pm$0.0 & 1.0$\pm$0.0 & \textbf{0.1520}$\pm$0.1089 & \textbf{0.1147}$\pm$0.0744 & \textbf{3.7941} & \textbf{4.1970} \\
    \bottomrule
    \end{tabular}  
\end{table*}

\textbf{Simulated data.} We also evaluated our approach with simulated data. The benefits of simulated data are two-fold: (i) we have ground truth values for all parameters to estimate how preferable the deployment and execution strategies are in HCs, and (ii) we can control compute capacity, locations, and bandwidth for example to study their impact on the performance of privacy-preserving distribution data processing. To do so, we selected a set of geographically distributed GPU nodes from participating universities and CSP's on-demanding cloud resources\footnote{Example of commercial Cloud Infrastructure offering GPU servers. \url{https://console2.fluidstack.io/virtual-machines} Retrieved 14 July 2023.}. 
We investigated the relative performance among 29 GPU servers based on TPU review data\footnote{GPU Database--TechPowerUp. \url{https://www.techpowerup.com/gpu-specs/} Retrieved 14 July 2023.} and tested over a remote virtual machine equipped with a Tesla T4 16GB GPU device to get the performance data. This model of GPU is sufficiently representative of the others used in HC.

We evaluated the workload's performance on a remote server equipped with a Tesla T4 16GB GPU device. The latency was measured by \textit{the processing time} (seconds), and the throughput was given by \textit{No. Of patches/latency} (patches per second). The observed maximum throughput $txt=200$ was set as our performance parameter for simulation experiments. 
We set various bandwidths with different geo-locations; for example, the network throughput from different cities was referred to as the fixed upload Internet speeds provided by SpeedTest by Ookla\footnote{Example of Amsterdam Fixed Speeds. \url{https://www.speedtest.net/performance/netherlands/north-holland/amsterdam} Retrieved 7 September 2023}. 

% \subsubsection{Evaluation Metrics}
% The mainstream work on privacy-preserving distributed processing uses information-theoretic metrics like average information gain, output utility, and lower bound on individual privacy since the information-theoretic security assumes a stronger adversary and is more efficient. Moreover, the Pareto frontiers are widely used for evaluating the solutions to MOO problems (See in section~\ref{sec:preliminar}). 

\subsection{Experiments}
We conducted two experiments. (i) Our first experiment addresses the privacy-preserving data splitting problem by comparing the baseline, i.e., average data splitting, with graph-based randomized data splitting extensions. We use one real data set containing 10 WSIs and key metrics (See in section~\ref{sec:preliminar}) to evaluate the robustness of the data splitting approaches. (ii) Our second experiment addresses the service deployment optimization problem formulated as the MOO by comparing the Pareto frontier or Pareto optimal solutions of our approach and that from evenly data splitting. We use the simulated dataset to evaluate the performance of the algorithms. We report the results of the proposed approach considering privacy, budget, and time constraints and Pareto dominance. 
% (iii) Our third experiment addresses the boundary analysis between centralized computing and our privacy-preserving distributed computing. 

\subsection{Implementation}
The algorithms and experiments were implemented and evaluated based on the DP case study with a virtual environment, in which we used Python 3.9.16, Pandas 1.5.3, Numpy 1.22.3, Numba 0.57.1, pareto\_2d, pareto\_3d, complemented with Pytorch 1.13.1 and NetworkX 3.0 libraries. To validate the average information gain, output utility, and individual privacy on lower bound, we ran real data experiments and used the \textit{SciPy} and \textit{sklearn} python libraries to calculate the Shannon entropy and estimate the normalized mutual information (NMI) with the measurements 
\textit{entropy}\footnote{Entropy. \url{https://docs.scipy.org/doc/scipy/reference/generated/scipy.stats.entropy.html} Retrieved 7 September 2023.} and \textit{normalized\_mutual\_info\_score}\footnote{Normalized mutual information. \url{https://scikit-learn.org/stable/modules/generated/sklearn.metrics.normalized_mutual_info_score.html} Retrieved 7 September 2023.}, respectively. 
% We have applied our proposed system model to a dataset described in Table.~\ref{tab: data}. 
% , as listed in Appendix Table~\ref{tab: infracloud}. 
During the experiment, 
% During the experimentation, since the resource nodes (i.e., GPU servers) are heterogeneous and the workloads are diverse, the time performance and cost vary for each computational task.
we use a trusted machine with Apple M1/Intel mode, 16GB of RAM, and 2TB Solid State USB Drive Storage to run the use case and simulation experiments.

\section{Results and Evaluations}\label{sec:results}

\subsection{Performance of Privacy Preservation}

Fig~\ref{fig: ppdatasplit} provides visual results of a WSI obtained by our proposed privacy-preserving data splitting. We first removed the background of the original data to generate a binary mask and then created some patches of size (224, 224, 3) from the WSI. Based on the coordinates of patches, we built a graph to split the patches into different patch-level sub-datasets and marked them using the `Viridis' color scale. Each color (e.g., yellow, green, dark green, blue, and dark blue) in the graph (the right picture) indicates a single data set that will be stored at an individual processing node, respectively. 

% As shown in the figure, there are five colors, i.e., the digital slide has been divided into five sub-datasets using our approach. Moreover, Table.~\ref{tab: dataset} lists the split data of the original data set (See in Table~\ref{tab: data}). The split data is unbalanced in different sizes and patch numbers based on the randomness of the splitting method. 

Table~\ref{tab: privacyeval} lists the measurement about the output utility, individual privacy (lower bound), and the average information gain on the real dataset in comparisons of the baseline and our approach. The average data splitting can still reach a perfect output utility combined with our data perturbation procedure. However, our approach can achieve lower mean individual privacy by increasing the uncertainty of split data sizes. That is, even though $\mathcal{N}-1$ nodes in the untrusted zone were corrupted, revealing private data to the adversaries, the amount of information about the private data learned by the adversaries over the remaining node of our approach is still less than the baseline method. Besides, our average information gain is bigger than that of the baseline. Hence, our privacy-preserving algorithm is more robust than the baseline method.

% Note that we also tested ICA-based attacks on each perturbed data set to reconstruct the original data. However, for each perturbed data, it cannot converge; that is to say, the ICA-based attack is not valid to the perturbed data. 

% \input{Tables/table7}

\subsection{Quality of the Pareto Fronts} 
% To answer the entire research question problem, we study whether our solution performs better than the baseline on service allocation optimization with heterogeneous untrusted cloud instances and different workloads over an HC. 

% \textcolor{blue}{Fig.~\ref{fig: paretores} shows the comparisons of several Pareto frontiers collected from various user preference settings, in which sub figures~\ref{fig: base-a},~\ref{fig: base-b}, and~\ref{fig: base-c} are baseline's results and ours are sub figures~\ref{fig: ours-a},~\ref{fig: ours-b}, and~\ref{fig: ours-c}. Each figure presents the feasible solutions, consisting of the non-Pareto optimal and Pareto optimal solutions of the service allocation optimization problem, and lists the first 6 out of all Pareto optimal solutions for clarity, presented as $\text{Pfdata}[0-5]$. In some cases, the Pareto optimal solutions contain diverse data points with all three dimensions that form a shell of a polygon covering all non-Pareto optimal values. Sometimes, the Pareto optimal solutions may only contain the same number of servers. Thus, the Pareto frontier performs a curve. Alternatively, there could be only one Pareto optimal value in the feasible solutions. }

\begin{figure}[!ht]
    \centering
    \includegraphics[width=0.48\textwidth]{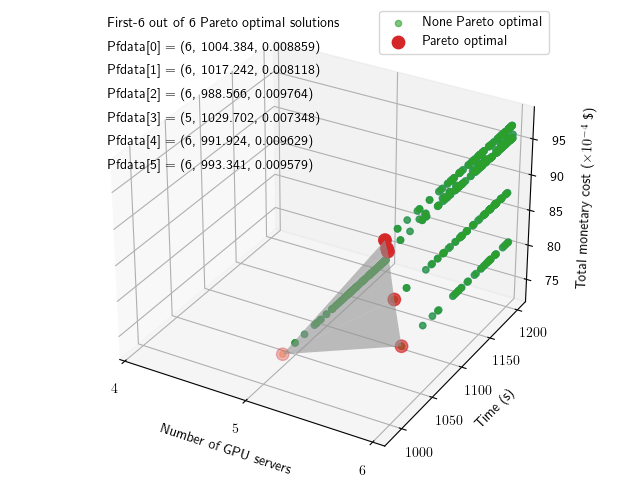}
    \caption{A Pareto front obtained by our approach with constraints, i.e., Nodes=8, Budget=0.01, Time Threshold=1200.}
    \label{fig: paretores}
\end{figure}

Fig.~\ref{fig: paretores} shows an example of the Pareto front obtained by our approach, in which it contains six Pareto optimal solutions $\text{Pfdata}[0-5]$ under user predefined constraints, i.e., the maximum number of untrusted processing nodes (No), budget, and time cost over untrusted nodes. The Pareto optimal solutions form a shell of a polygon covering all non-Pareto optimal values. Each solution is equivalent in the Pareto set. Table~\ref{tab: pfeval} lists the comparisons of several Pareto optimal solutions collected from various user preference settings, such as ``Normal'', ``Very Small Number (VSN)'', ``Very Tight Budget (VTB)'', ``Very Short Time (VST)''. For instance, Fig.~\ref{fig: paretores} is a mapping of VTB user preference. 

\begin{table*}[!ht]
    \centering
    \caption{Comparisons of our approach with the baseline method in terms of the Pareto optimal-based service allocation.}
    \label{tab: pfeval}
    \begin{tabular}{cccccccccc}
    \toprule
        \multirow{2}*{\textbf{User Preferences}} & \multirow{2}*{\textbf{Nodes}} & \multirow{2}*{\textbf{Budget}} & \multirow{2}*{\textbf{Time Threshold}} & \multicolumn{3}{c}{\textbf{Baseline }} & \multicolumn{3}{c}{\textbf{Our Approach}} \\ \cline{5-10}
        \textbf{} & \textbf{} & \textbf{} & \textbf{} & \textbf{$f_1 (\downarrow)$} & \textbf{$f_2 (\downarrow)$} & \textbf{$f_3 (\downarrow)$} & \textbf{$f_1 (\downarrow)$} & \textbf{$f_2 (\downarrow)$} & \textbf{$f_3 (\downarrow)$} \\
    \midrule
       Normal  & 8 & 0.1 & 800 & 7 & 0.084264 & 797.981 & \textbf{6} & \textbf{0.033261} & \textbf{743.688} \\
       Very Small No. Of nodes (VSN) & 5 & 0.2 & 1200 & 5 & 0.049659 & 1060.841 & 5 & \textbf{0.007348} & \textbf{1029.702} \\
       Very Tight Budget (VTB)  & 8 & 0.01 & 1200 & N/A & N/A & N/A & 6 & 0.008859 & 1004.384 \\
       Very Short Time (VST)   & 8 & 0.2 & 560 & 7 & 0.118831 & 535.345 & N/A & N/A & N/A \\       
       
    \bottomrule
    \end{tabular}  
\end{table*}

As listed in Table~\ref{tab: pfeval}, our solution can entirely dominate the baseline's solution for usual user preference since it has fewer untrusted nodes, less total monetary cost, and fewer time costs. Besides, for the VSN preference, our approach performs better than the baseline regarding the monetary and time costs. Specifically, our approach has been ahead of baseline by 60.53\% and 6.8\% (See ``Normal''), 85.2\% and 2.94\% (See ``VSN'') in terms of the total financial cost ($f_2$) and time cost ($f_3$), respectively. Furthermore, our approach can also support a very tight budget in the inapplicable cases of the baseline (See ``VTB''). However, assuming the application is very time-critical (See ``VST''), our approach may not be available.

\subsection{Limitations and Scope}
Even though we have successfully obtained the Pareto optimal based service allocation results for our research problem, we encountered certain limitations and setbacks. First, the communication overhead has been simply predicted by network throughput and request size. This setting limits the network uncertainties and complexities. There do exist communication complexities, such as uncertainty of worldwide network latency, diverse bandwidth under SLA constraints, and firewall configurations. Besides, for big data processing, there will be several performance bottlenecks regarding the network throughput between the trusted zone and untrusted zone. Further studies should consider such complexities to obtain a more accurate representation of realistic applications. Second, the performance benchmark is still missing, even though we estimated the time using the inference processing time, throughput, and relative performance data. However, since our focus is mainly on the MOO decision-making based on input data, the quality of performance benchmark data will determine that of decision-making. The results would be more reliable for users if we could obtain a powerful benchmarking data set. Third, our solution is not applicable for the cases of very time-critical in the simulation experiments. The main reason could be the unbalanced data size increases the heterogeneity of the workloads compared to homogeneous workloads of the baseline.

\section{Conclusion}\label{sec:conclusion}
In this paper, we focus on customizing a virtual infrastructure and scheduling the large image processing workflow in a distributed and privacy-preserving way. We formulated the problem statement as a privacy-, cost-, and time-aware service allocation optimization problem in HCs for large medical image processing. To address it, we proposed a two-phase approach: (1) a novel privacy mechanism design was implemented to preserve sensitive attributes from the architecture structure and perturbative masking perspectives, and (2) a greedy Pareto front-based algorithm to solve the MOO service allocation under privacy, budget, and time requirements. Our study bridges the gap in this field and demonstrates the problem-solving and evaluation process. At the key aspects of privacy evaluation, our approach outperforms the baseline method regarding the average individual privacy at the lower bound and average information gain. Moreover, our Pareto optimal solutions are superior to the baseline on very tight budget cases. In the majority of cases, they can entirely dominate the baseline solutions. In the future, we will improve the quality of solutions and integrate our approach as a component of a Jupyter-based virtual research environment.

\section*{Author Contributions}
Yuandou Wang: Conceptualization of this study, Methodology, Investigation, Writing - Original draft preparation, Writing - Review \& Editing. Neel Kanwal: Provide the preprocessing pipeline for detecting artifacts in whole slide images that can be ran over a single GPU server, Investigate the related work on privacy-preserving techniques and image processing, WSI Data collection, Writing - Review \& Editing. Engan Kjersti: Writing - Review \& Editing, Supervision. Chunming Rong: Conceptualization, Writing - Review. Paola Grosso: Conceptualization, Writing - Review \& Editing, Supervision. Zhiming Zhao: Conceptualization, Writing - Review \& Editing, Supervision, Project administration, Funding acquisition.

% \section*{Acknowledgment}
% This work has been funded by the European Union project CLARIFY (860627), $\text{ENVRI}^\text{FAIR}$ (824068), BlueCloud-2026 (101094227) and LifeWatch ERIC. 

\bibliographystyle{IEEEtran}
\bibliography{refs.bib}

\end{document}